\documentclass[11pt,a4paper]{article}
\pdfoutput=1
\usepackage{amsmath,amssymb,tikz}
\usepackage{graphicx}
 \allowdisplaybreaks
\def\be{\begin{equation}}
\def\ee{\end{equation}}
\def\ba#1\ea{\begin{align}#1\end{align}}
\def\bg#1\eg{\begin{gather}#1\end{gather}}
\def\bm#1\em{\begin{multline}#1\end{multline}}
\def\bmd#1\emd{\begin{multlined}#1\end{multlined}}

\def\la{\label}

\def\({\left(}
\def\){\right)}
\def\[{\left[}
\def\]{\right]}

\def \be {\begin{equation}}
\def \ee {\end{equation}}
\def \ba {\begin{array}}
\def \ea {\end{array}}
\def \bea{\begin{eqnarray}}
\def \eea{\end{eqnarray}}

\def \la {\leftarrow}
\def \ra {\rightarrow}

\def\bea{\begin{eqnarray}}
\def\eea{\end{eqnarray}}

\newcommand{\bit}{\begin{itemize}}  \newcommand{\eit}{\end{itemize}}
\newcommand{\ben}{\begin{enumerate}}  \newcommand{\een}{\end{enumerate}}

\def\la{\langle}
\def\ra{\rangle}

\long\def\symbolfootnote[#1]#2{\begingroup%
\def\thefootnote{\fnsymbol{footnote}}\footnote[#1]{#2}\endgroup}
\newcommand{\sysu}{{\it School of Physics and Astronomy, Sun Yat-Sen University, 2 Daxue Road, Zhuhai 519082, China}}
\begin{document}
%%%%%%%%%%%%%%%%%%%%%%%%%%%%%%%%%%%%%%%%%%%
\thispagestyle{empty}
%%%%%%%%%%%%%%%%%%%%%%%%%%%%%%%%%%%%%%%%%%%
%%%%%%%%%%%%%%%%%%%%%%%%%%%%%%%%%%%%%%%%%%%
%\begin{flushright}
%\hfill{AEI-2015-xxx}
%\hfill{ NCTS-TH/1702}
%\end{flushright}
%%%%%%%%%%%%%%%%%%%%%%%%%%%%%%%%%%%%%%%%%%%
\begin{center}

~\vspace{20pt}

{\Large\bf Holographic Anomalous Chiral Current Near a Boundary}

\vspace{25pt}

Rong-Xin Miao ${}$\symbolfootnote[1]{Email:~\sf
  miaorx@mail.sysu.edu.cn}, Yu-Qian Zeng

\vspace{10pt}${}$\sysu

\vspace{2cm}

\begin{abstract}

Due to the Weyl anomaly, an axial vector field produces novel anomalous chiral currents in spacetime with boundaries. Remarkably, the chiral current is not invariant under the gauge transformation of axial vector fields. As a result, more potential terms appear, and the chiral current becomes larger than the electric current near the boundary. This paper investigates the anomalous chiral current in AdS/BCFT. We find that the minimal holographic model of the axial vector field cannot produce the expected chiral current. To resolve this problem, we propose adding a suitable boundary action. Furthermore, we notice a similar situation for holographic condensation. Finally, we obtain the shape dependence of holographic chiral current and verify that it agrees with the field-theoretical result. 
\end{abstract}

\end{center}

%%%%%%%%%%%%%%%%%%%%%%%%%%%%%%%
\newpage
\setcounter{footnote}{0}
\setcounter{page}{1}
%%%%%%%%%%%%%%%%%%%%%%%%%%%%%%%

\tableofcontents
%%%%%%%%%%%%%%%%%%%%%%%%%%%%%%%

\section{Introduction}

Anomaly-induced transports are novel phenomena with wide applications in condensed matter, quantum field theory, and cosmology \cite{review}. The famous examples include the chiral magnetic effect (CME)  \cite{Vilenkin:1995um,
Vilenkin:1980fu, Giovannini:1997eg, alekseev, Fukushima:2012vr} and chiral vortical effect (CVE) \cite{Kharzeev:2007tn,Erdmenger:2008rm,Banerjee:2008th,Son:2009tf,Landsteiner:2011cp,Golkar:2012kb,Jensen:2012kj}, which originated from the chiral anomaly. See also \cite{Chernodub:2016lbo, Chernodub:2017jcp,Chu:2018fpx,Chu:2019rod,Miao:2017aba,Miao:2018dvm,Chernodub:2018ihb,Chernodub:2019blw,Ambrus:2019khr,Zheng:2019xeu,Hu:2020puq,Kawaguchi:2020kce,Kurkov:2020jet,Kurkov:2018pjw,Fialkovsky:2019rum} for related works.  Recently, it has been found that the Weyl anomaly \cite{Duff:1993wm} leads to novel anomalous currents \cite{Chu:2018ksb,Chu:2018ntx}, Fermi condensations \cite{Chu:2020mwx,Chu:2020gwq}  and chiral currents \cite{Chu:2022bhj} in spacetime with boundaries. Remarkably, the Weyl-anomaly-induced transports take universal forms near the boundary, which apply to not only conformal field theory but also the general quantum field theory because the Weyl anomaly is well-defined for general quantum field theory \cite{Duff:1993wm}. 

Take free Dirac field as an example
\begin{eqnarray}\label{Diracaction}
  S=\int_M \sqrt{-g}\bar{\psi} i \gamma^{\mu} \big(\nabla_{\mu}-i V_{\mu}-i \gamma_5 A_{\mu} \big) \psi,
\end{eqnarray}
where $V_{\mu}$ and $A_{\mu}$ are  vectors and axial vectors, respectively. We impose the general chiral bag boundary conditions \cite{Chodos:1974je,Chodos:1974pn}  
\begin{eqnarray} \label{bagBC}
(1+i e^{i \theta \gamma_5} \gamma^n)\psi|_{\partial M}=0
\end{eqnarray}
where $n$ denotes the normal direction and $\theta$ is a constant. At the leading order near the boundary, various anomalous transports near the boundary are given by  \cite{Chu:2018ksb,Chu:2018ntx,Chu:2020mwx,Chu:2020gwq}
\begin{eqnarray}\label{current}
&&{\text{current}}:\ \ \ \la J_V^{\mu} \ra=\frac{1}{6\pi^2}\frac{ n_{\nu}H^{\nu\mu}}{x} + O(\ln x), \\ \label{condensation}
&&{\text{condensation}}:\  \la\bar{\psi}\psi\ra= \frac{\cos\theta}{4\pi^2}\frac{1}{x^3}+ O( 1/x^2), \\ \label{chiral current}
&&{\text{chiral current}}:\ \ \ \la J_A^{\mu} \ra=\frac{-1}{6\pi^2}\frac{h^{\mu\nu}A_{\nu}}{x^2}+ O(1/x), 
\end{eqnarray}
where $H=dV$ denotes the field strength of vectors, $x$ is the distance to the boundary, 
$n_{\nu}$ is the inward-pointing normal vector and $h_{\mu\nu}$ is the induced metric on the boundary. Note that we take signature $(-1,1,1,1)$ in this paper, which is different from  \cite{Chu:2022bhj}. Note also that the above results apply to $x>\epsilon$, where $\epsilon$ is a cut-off. There are also boundary contributions at $x=\epsilon$ to the current and chiral current, which make the total transports finite  \cite{Chu:2018ksb}.

This paper studies the anomalous chiral current in AdS/BCFT \cite{Takayanagi:2011zk}. We find that the minimal holographic model of axial vectors cannot reproduce the expected chiral current (\ref{chiral current}). To resolve this problem, we propose adding a relevant boundary term, which is similar to the case of holographic condensation. We also investigate the shape dependence of holographic chiral current and verify that it obeys the non-trivial constraint from the Weyl anomaly.

The paper is organized as follows. To warm up, we first discuss holographic condensation in section 2. We find that an additional boundary action is necessary in order to derive the expected condensation. In section 3, we study a non-minimal holographic model of the axial vector field and derive the desired anomalous chiral current. In section 4, we investigate the shape dependence of holographic chiral current. Finally, we conclude with some open questions in section 5.

\section{Holographic condensation}

To warm up, we first study holographic condensation in this section. Due to the novel boundary effects, the vacuum expectation of a scalar operator is non-zero 
\cite{Billo:2016cpy}
\begin{eqnarray} \label{BCFTcondensation}
\la O\ra \sim \frac{1}{x^{\Delta}},
\end{eqnarray}
where $\Delta$ is the conformal dimension, and $x$ is the distance to the boundary. For $O=\bar{\psi}\psi$ 
and $\Delta=3$, (\ref{BCFTcondensation}) becomes the Fermi condensation (\ref{condensation}). It is found in \cite{Fujita:2011fp} that the minimal holographic model cannot recover the result (\ref{BCFTcondensation}). Instead, one has to add a boundary action on the end-of-the-world brane $Q$,
\begin{eqnarray} \label{action}
I=\int_N d^{d+1}x\sqrt{|g|}
\left(R+d(d-1)-\frac{1}{2}\left(\nabla^{\mu}
\phi\nabla_{\mu} \phi+ m^2 \phi^2\right)
\right)+2\int_Q dx^d\sqrt{|\gamma|} \left(K-T +\frac{\xi}{2}
\phi \right),
\end{eqnarray}
where $K$ is the extrinsic curvature, $T=(d-1)\tanh\rho$ is the brane tension
% v10
\footnote{Note that $\rho$ is a constant, which is equal to $\rho^*$ of  \cite{Takayanagi:2011zk}. }
, $\xi$ is a parameter 
that plays an essential role in deriving condensation (\ref{BCFTcondensation}).  Above we have set $16 \pi G_N=1$ and AdS radius $l=1$.  
% v10
For simplicity, following \cite{Takayanagi:2011zk}, we do not show the boundary terms on the AdS boundary, which take the same forms as those in AdS/CFT. 
Following \cite{Takayanagi:2011zk}, we choose 
the Neumann boundary
condition (NBC) on $Q$
\begin{eqnarray} \label{NBCmetric}
&&K_{ij}-(K-T+\frac{\xi}{2} \phi)\gamma_{ij} =0,\\ &&
% choose $n^{\mu}$ to be inward-pointing, as a result $n^{\mu}$  change sign
  n^{\mu}\nabla_{\mu} \phi+
  \xi=0, \label{NBCscalar}
\end{eqnarray}
where $n^{\mu}$ is the inward-pointing normal vector on $Q$.
See \cite{Miao:2018qkc,Chu:2021mvq} for other choices of  boundary conditions. For simplicity, we focus on the probe limit with the following bulk metric and embedding function of brane $Q$ 
\begin{eqnarray} \label{AdSmetric}
&& {\text{bulk metric}}:\ \ ds^2=\frac{dz^2+dx^2+\eta_{ab}dy^a dy^b}{z^2},\\ \label{Q}
&& {\text{Q}}:\ \ x=-\sinh(\rho) z,
\end{eqnarray}
% v10
where $\eta_{ab}=\text{diag}(-1,1,...,1)$.
Following \cite{Chu:2020gwq}, we take the following ansatz of scalar field
\begin{eqnarray} \label{scalar}
\phi=f(\frac{z}{x}).
\end{eqnarray}
Substituting (\ref{scalar}) into the equation of motion (EOM) of scalar and denoting $s=z/x$, we obtain
\begin{eqnarray} \label{EOMscalar}
s \left( s \left(s^2+1\right) f''(s)+\left(2 s^2+1-d\right) f'(s)\right)-m^2 f(s)=0,
\end{eqnarray}
The above equation can be solved as
\begin{eqnarray} \label{solutionscalar}
f(s)=&&c_1 s^{\Delta } \, _2F_1\left(\frac{\Delta }{2},\frac{\Delta +1}{2};1-\frac{d}{2}+\Delta;-s^2\right)\nonumber\\
&+&c_2 s^{d-\Delta } \, _2F_1\left(\frac{d-\Delta }{2},\frac{d-\Delta +1}{2};1+\frac{d}{2}-\Delta;-s^2\right),
\end{eqnarray}
which yields the expected asymptotic solution near the AdS boundary
\begin{eqnarray} \label{AdSbdyscalar}
 \phi=z^{d-\Delta}\phi_0 +z^{\Delta}\frac{1}{2\Delta -d} \la O\ra, \quad z\to 0,
\end{eqnarray}
where $\Delta=\frac{d}{2}+\sqrt{(\frac{d}{2})^2+m^2}$ is the conformal dimension of the operator
$O$ dual to $\phi$, and $\phi_0$ is the background scalar coupled to CFTs on the AdS boundary.  Following \cite{Chu:2020gwq}, we choose $\phi_0=0$ on the AdS boundary, which gives $c_2=0$.
Now imposing NBC (\ref{NBCscalar}) on the brane, we fix the integral constant
\begin{eqnarray} \label{c1scalar}
	c_1=\frac{\xi  \tanh (\rho ) (-\text{csch}(\rho ))^{-\Delta }/\Delta}{\frac{(\Delta +1) \text{csch}^2(\rho ) \, _2F_1\left(\frac{\Delta +2}{2},\frac{\Delta +3}{2};-\frac{d}{2}+\Delta +2;-\text{csch}^2(\rho )\right)}{-d+2 \Delta +2}-\, _2F_1\left(\frac{\Delta }{2},\frac{\Delta +1}{2};-\frac{d}{2}+\Delta +1;-\text{csch}^2(\rho )\right)}.
\end{eqnarray}
From (\ref{solutionscalar}-\ref{c1scalar}), we finally derive the holographic condensation
\begin{eqnarray} \label{holographic condensation}
\la O\ra =(2\Delta-d) \frac{c_1}{x^{\Delta}} \sim  \frac{\xi}{x^{\Delta}},
\end{eqnarray}
which agrees with field-theoretical result (\ref{BCFTcondensation}). Note that $\la O\ra$ vanishes in the minimal holographic model with $\xi=0$. 

Let us make some comments. First, to derive the expected condensation (\ref{BCFTcondensation}), one has to consider a non-minimal holographic model with an additional boundary term on the brane. As we will discuss in the next section, the case of chiral current is similar. Second, the results of this section are not
new. The special case with $d=4$ and $\Delta=3$ is already studied in \cite{Chu:2020gwq}. The general case is also studied in \cite{Fujita:2011fp}. Here we apply a different method developed in \cite{Miao:2017aba}, which is simpler than the initial method of \cite{Fujita:2011fp}. 

\section{Holographic chiral current}

In this section, we discuss the holographic chiral current in a flat half space. For our purpose, we focus on the most interesting case $d=4$. Similar to 
holographic condensation, we consider the non-minimal model of axial vectors with a boundary action on $Q$
\begin{eqnarray} \label{vectoraction}
I=\int_N d^{5}x\sqrt{|g|}
\left(-\frac{1}{4}\mathcal{F}_{\mu\nu}\mathcal{F}^{\mu\nu}+\frac{\kappa}{2}\epsilon^{\mu\nu\rho\sigma\alpha} \mathcal{A}_{\mu}\mathcal{F}_{\nu\rho}\mathcal{F}_{\sigma\alpha}\right)+\int_Q dx^4\sqrt{|\gamma|} \left(\frac{ \lambda}{2} \gamma^{\mu\nu}\mathcal{A}_{\mu}\mathcal{A}_{\nu}\right),
\end{eqnarray}
where $\mathcal{F}=d\mathcal{A}$,
$\lambda$ is a constant, and $\gamma_{\mu\nu}$ is the induced metric on the brane. Note that the above action is  not invariant under the gauge transformation 
of axial vectors
$\delta \mathcal{A}_{\mu}=\partial_{\mu}\alpha$. 
This is necessary to derive the non-gauge-invariant chiral current (\ref{chiral current}).  From (\ref{vectoraction}), we derive the EOM
\begin{eqnarray} \label{axialvectorEOM}
\nabla_{\mu}\mathcal{F}^{\mu\nu}+\frac{3\kappa}{2}\epsilon^{\nu\rho\sigma\alpha\beta} \mathcal{F}_{\rho\sigma}\mathcal{F}_{\alpha\beta}=0
\end{eqnarray} 
and NBC on the brane $Q$
% note that n_{\mu} is inward-pointing 
\begin{eqnarray} \label{axialvectorNBC}
n_{\mu}(\mathcal{F}^{\mu\nu}-2\kappa \epsilon^{\mu\nu \rho\alpha\beta} \mathcal{A}_{\rho}\mathcal{F}_{\alpha\beta})+\lambda \mathcal{A}_{\mu}
 \gamma^{\mu\nu}=0.
 \end{eqnarray} 
 
 Similar to the above section, we focus on the probe limit (\ref{AdSmetric},\ref{Q}).  We take the following ansatz %v2
 of bulk axial vectors 
 % v10
 \footnote{Note that one cannot modify the ansatz (\ref{axialvector}) to give the correct chiral current without the boundary term in action (\ref{vectoraction}). See the conclusion section for reasons. }
 \begin{eqnarray} \label{axialvector}
\mathcal{A}_{z}=\mathcal{A}_{x}=0, \ \mathcal{A}_{a}= A_a f_0(\frac{z}{x}),
 \end{eqnarray} 
where $a$ denotes the direction parallel to the boundary $x=0$, $A_a$ is a constant vector and we choose the Dirichlet boundary condition (DBC) on the AdS boundary 
 \begin{eqnarray} \label{axialvectorDBC}
f_0(0)=1,
 \end{eqnarray} 
 so that $A_a $ of (\ref{axialvector}) is the axial vector field coupled to CFTs on the AdS boundary.  
 Substituting (\ref{axialvector}) into EOM (\ref{axialvectorEOM}), we obtain 
 \begin{eqnarray} \label{axialvectorEOMf0}
 s \left(s^2+1\right) f_0''(s)+\left(2 s^2-1\right) f_0'(s)=0,
  \end{eqnarray} 
  which can be solved as 
  \begin{eqnarray} \label{axialvectorEOMf0}
  f_0(s)=1+b_1-\frac{b_1}{\sqrt{s^2+1}},
 \end{eqnarray}
 where we have used DBC (\ref{axialvectorDBC}). Imposing NBC (\ref{axialvectorNBC}) on the brane, we derive the integral constant
  \begin{eqnarray} \label{axialvectorb1}
  b_1=\frac{\lambda  \cosh (\rho ) \coth (\rho )}{\text{csch}(\rho )-\lambda  \cosh (\rho ) (\coth (\rho )+1)}.
 \end{eqnarray}
 Substituting (\ref{axialvector},\ref{axialvectorEOMf0}) into the holographic formula of chiral current \cite{Gynther:2010ed}
  \begin{eqnarray} \label{axialvectorholoJ}
 \sqrt{|h|} J_A^{\nu}=\lim_{z\to 0} \sqrt{|g|}(\mathcal{F}^{z\nu}-2\kappa \epsilon^{z\nu \rho\alpha\beta} \mathcal{A}_{\rho}\mathcal{F}_{\alpha\beta}),
 \end{eqnarray}
 we finally obtain the holographic chiral current 
   \begin{eqnarray} \label{holochiralcurrent }
  J_{A \ a}=b_1 \frac{A_a}{x^2},
 \end{eqnarray}
 which takes the expected form (\ref{chiral current}). 
 % v10
Note that the second term on the right-hand side of (\ref{axialvectorholoJ}) is of order $O(A^2)$ and vanishes strictly for the solution (\ref{axialvector},\ref{axialvectorEOMf0}). Note also that (\ref{holochiralcurrent }) for holographic CFTs needs not to be equal to (\ref{chiral current}) for free Dirac fields.  Above $g_{\mu\nu}$ is given by the metric (\ref{AdSmetric}) with $d=4$. Thus we have $\sqrt{|g|}=1/z^5$. $h_{ij}$ denotes the metric of BCFT. For our present case, BCFT lives in a flat half space with $\sqrt{|h|} =1$.  Now we finish the holographic derivation of the chiral current in a flat half space.

\section{Shape dependence of holographic chiral current}

In this section, we investigate the shape dependence of holographic chiral current. The shape of the boundary is described by the extrinsic curvature $k_{ab}$, which affects the chiral current at sub-leading order \cite{Chu:2022bhj}
  \begin{eqnarray} \label{sect4: linear order}
J_A \sim \frac{A}{x^2}+ \frac{k A }{x}+...
 \end{eqnarray}
For simplicity, we focus on the results in the linear order of the extrinsic curvature $k$ and the background axial vector field $A$.    Besides, we consider constant axial vector fields $A=\lim_{z\to 0}\mathcal{A}$ 
and extrinsic curvatures $k$ on the AdS boundary.
 
Note that the back-reaction of axial vector fields to the geometry is of order $O(A)^2$. Therefore, at the linear order of $O(A)$, we can ignore the back-reaction safely.  For a non-flat boundary, the bulk metric 
and the embedding function of $Q$ are given by \cite{Miao:2017aba}
\begin{eqnarray}\label{bulkmetriclimit}
ds^2=\frac{1}{z^2}\Big{[} dz^2+ dx^2 +\big(\delta_{ab}-2x\epsilon \bar{k}_{ab}
    f(\frac{z}{x})-2x\epsilon \frac{k}{3} \delta_{ab}  \big)dy^a dy^b\Big{]}+O(\epsilon^2),
\end{eqnarray}
and
\begin{eqnarray}\label{4dQlimit}
x=-\sinh(\rho) z+\epsilon \frac{k \cosh ^2\rho }{6} z^2 +O(\epsilon^2),
\end{eqnarray}
where $k_{ab}=\text{diag}(k_1,k_2,k_3)$ is the extrinsic curvature, $k$ is the trace and $\bar{k}_{ab}$ is the traceless part of the extrinsic curvature, 
% v10 $\epsilon$ denotes the order of $k$, 
$\epsilon$ keeps track of the order of $k$ and should be set to 1 in the end, 
and $f(s)$ is given by 
\begin{eqnarray}\label{flimit}
&&f(s)=1+2 \alpha _1-\frac{\alpha _1
    \left(s^2+2\right)}{\sqrt{s^2+1}},\\ &&\alpha_1=\frac{-1}{2 (1+\tanh
    \rho)}. \label{sect4:a1}
\end{eqnarray}
We take the following ansatz of bulk axial vectors 
 \begin{eqnarray} \label{sect4:axialvector}
\mathcal{A}_{z}=\mathcal{A}_{x}=0, \ \mathcal{A}_{a}= \epsilon_1 A_a  \big(  f_0(\frac{z}{x}) +\epsilon\  x f_{1a}(\frac{z}{x})  \big),
 \end{eqnarray} 
% v10 where $\epsilon_1$ denotes the order of $A$, 
where $\epsilon_1$ keeps track of the order of $A$, 
and we choose the DBC 
 \begin{eqnarray} \label{sect4:axialvectorDBC}
f_0(0)=1, \  f_{1a}(0)=0
 \end{eqnarray} 
so that $A_a=\lim_{z\to 0} \mathcal{A}_{a}$ is a constant axial vector on the AdS boundary.  Following the approach of section 3, we solve EOM (\ref{axialvectorEOM}) together with BCs (\ref{axialvectorNBC},\ref{sect4:axialvectorDBC}) and obtain at order $O(\epsilon \epsilon_1)$
 \begin{eqnarray} \label{sect4:f1a}
f_{1a}(s)&=&\bar{k}_{aa}\Big[b_1 \left(\frac{2}{3} \alpha _1 \left(\frac{3}{\sqrt{s^2+1}}-\frac{1}{s^2+1}-2\right)+\frac{1}{\sqrt{s^2+1}}-1\right)+d_1 \left(\sqrt{s^2+1}-1\right)\Big]\nonumber\\
&+& k \Big[ \frac{1}{6} b_1 \left(1-\frac{1}{\sqrt{s^2+1}}\right)+d_2 \left(\sqrt{s^2+1}-1\right) \Big],
 \end{eqnarray} 
where the repeat index of $\bar{k}_{aa}$ does not sum, $b_1$, $\alpha_1$ are given by (\ref{axialvectorb1},\ref{sect4:a1}), and $d_1, d_2$ are given by
 \begin{eqnarray} \label{sect4:d1}
d_1=\frac{\lambda  \cosh (\rho ) ((\lambda -11) \sinh (\rho )+(\lambda +1) \sinh (3 \rho )-(2 \lambda +3) \cosh (\rho )+(2 \lambda -1) \cosh (3 \rho ))}{3 \left(\lambda  e^{2 \rho }+\lambda -2\right)^2},
 \end{eqnarray} 
 and
  \begin{eqnarray} \label{sect4:d2}
d_2=\frac{\lambda  \cosh ^2(\rho )}{-3 \lambda  e^{2 \rho }-3 \lambda +6}.
 \end{eqnarray} 
 Substituting the solution (\ref{sect4:axialvector}) into the holographic formula (\ref{axialvectorholoJ}) with $\sqrt{|h|}=1-x k+O(x^2)$, we obtain the shape dependence of holographic chiral current
   \begin{eqnarray} \label{sect4:holochiralcurrent}
  J_{A\ a}=b_1 \frac{A_a}{x^2}+ \frac{ b_2 \bar{k}_{a}^{\ b} A_b+b_3 k A_a}{x}+O(x^0,\ln x),
 \end{eqnarray}
 where 
 \begin{eqnarray} \label{sect4:b2}
 b_2=\frac{6 d_1-2 \left(2 \alpha _1+3\right) b_1}{6}=\frac{\lambda  \cosh (\rho ) ((2 \lambda -3) \cosh (\rho )+3 \sinh (\rho ))}{3 ((\lambda -1) \cosh (\rho )+\sinh (\rho ))^2},
 \end{eqnarray}
 and 
  \begin{eqnarray} \label{sect4:b3}
 b_3=\frac{1}{6} \left(b_1+6 d_2\right)=-\frac{2 \lambda  \cosh ^2(\rho )}{3 \left(\lambda  e^{2 \rho }+\lambda -2\right)}.
 \end{eqnarray}
 Now we finish the derivation of the shape dependence of holographic chiral current. Comparing (\ref{sect4:holochiralcurrent}) with the field-theoretical result (\ref{chiralcurrentJa}), we get a constraint
  \begin{eqnarray} \label{sect4:b1b3}
 b_3=\frac{1}{3} b_1. 
 \end{eqnarray}
 From (\ref{axialvectorb1}) and (\ref{sect4:b3}), we verify that (\ref{sect4:b1b3}) is indeed satisfied. This is a nontrivial check of our results.

\section{Conclusions and Discussions}

This paper investigates the holographic anomalous chiral current near a boundary. We find that one has to consider a non-minimal holographic model of axial vectors in order to produce the non-gauge invariant chiral current $J_A \sim A/x^2$. The reasons are as follows. The Chern-Simons term in the minimal model is of order $O(\mathcal{A}^3)$, which is irrelevant to the chiral current linear in $O(A)$. Ignoring the Chern-Simons term at leading order, the minimal action 
% v10 $F_A^2$ 
$\mathcal{F}^2_{\mathcal{A}}$
becomes gauge invariant and cannot give a non-gauge invariant chiral current. To recover the expected result $J_A \sim A/x^2$, one must add an action of order $O(\mathcal{A}^2)$ so that the EOM or BC is of order $O(\mathcal{A})$. One cannot add a bulk action $\mathcal{A}^2$ (affect EOM), since it would make the axial vector $\mathcal{A}$ massive and change the conformal dimension of dual chiral current $J_A$. The only choice is to add a boundary action $\int_Q \mathcal{A}^2$ (affect BC) on the brane. In this way, we derive the expected chiral current $J_A \sim A/x^2$. We notice that one also considers a non-minimal holographic model of scalars in order to derive the correct condensation. Finally, we work out the shape dependence of 
holographic chiral current and verify that it agrees with the results of field theories. It is interesting to generalize the discussions of this paper to Chern-Simons gravity and explore the corresponding one-point function of stress tensors. We hope we can address this problem in the future.

\section*{Acknowledgements}
We thank C. S. Chu for helpful discussions and comments. 
R. X. Miao acknowledges the supports from National Natural Science Foundation of China (No. 11905297) and Guangdong Basic and Applied Basic Research Foundation (No.2020A1515010900).

\appendix

\section{Chiral current from Weyl anomaly}

% v10
In this appendix, we focus on the conformal field theory in four dimensions instead of the holographic theory in five dimensions. 
% v11
In general, there are boundary contributions to the Weyl anomaly. See \cite{Herzog:2015ioa,Fursaev:2015wpa} for example. 
According to \cite{Chu:2022bhj}, the Weyl anomaly induced by a background axial vector takes the following form 
 \begin{eqnarray} \label{anomaly}
  \mathcal{A}=-
  \int_M dx^4\sqrt{-g} a_1 F_{\mu\nu}F^{\mu\nu}-\int_{\partial M}dx^3\sqrt{-h}[a_2(B_1-B_2)+ a_3\bar{k}_{\mu\nu}A^{\mu}A^{\nu}],
\end{eqnarray}
where $a_i$ are central charges and 
 \begin{eqnarray} \label{B1}
&& B_1=\frac{2}{3}k(n^{\mu}n^{\nu}-h^{\mu\nu})A_{\mu}A_{\nu}
+A_n \nabla_{\mu}A^{\mu}+2A_{\mu}h^{\mu\nu}\nabla_n A_{\nu},\\
&&B_2=n^{\mu}A^{\nu}\nabla_{\nu}A_{\mu}-\frac{1}{3}k A_{\mu}A^{\mu}.
\end{eqnarray}
Note that the notation of this paper is different from that of \cite{Chu:2022bhj}. In this paper, the metric signature is $(-1,1,1,1)$, $n_{\mu}$ is inward pointing,
% v10
$h_{\mu\nu}$ is the induced metric on the boundary, 
$k_{\mu\nu}=-h^{\alpha}_{\mu}h^{\beta}_{\nu}\nabla_{\alpha}n_{\beta}$ is the extrinsic curvature and $\bar{k}_{\mu\nu}$ is its traceless part. Although $B_1$ and $B_2$ are both Weyl invariant, only the combination $(B_1-B_2)$ can appear in Weyl anomaly. Otherwise, the chiral current derived from Weyl anomaly would have the normal component $J^{\mu}\sim n^{\mu}A_n/x^2$, which means that it cannot satisfy the conservation law $\nabla_{\mu} J^{\mu}\sim A_n/x^3 \ne O(1)$.  
% v11
According to \cite{Chu:2018ksb,Chu:2022bhj}, we have the  ``integrability'' relation
\begin{eqnarray} \label{key}
  (\delta \mathcal{A})_{\partial M_{\epsilon}}=\left(\int_{M_{\epsilon}} dx^4 \sqrt{-g} \ J_A^{\mu}
  \delta A_{\mu} \right)_{\log\frac{1}{\epsilon}},
\end{eqnarray}
where $\epsilon$ is the cutoff of $x$, i.e., the distance to the boundary,  $M_\epsilon$ is defined by the region $x\ge\epsilon$, $()_{\log\frac{1}{\epsilon}}$ means the coefficient of $\log\frac{1}{\epsilon}$, which comes from the $\frac{1}{x}$ term in the integral $\sqrt{-g} \ J_A^{\mu} \delta A_{\mu}$ since we have $\int_{\epsilon}^\infty dx \frac{1}{x} \sim \log\frac{1}{\epsilon}$. Please see \cite{Chu:2018ksb} for the proof of the ``integrability'' relation for the current. The generalization to the chiral current (\ref{key}) is straightforward. By applying (\ref{key}), one derives the chiral current \cite{Chu:2022bhj}
\begin{eqnarray}
  \label{chiralcurrentJa}
  &&J^{a}_A= \frac{ -2 a_2 A^a}{ x^2}+\frac{2(a_2-a_3)\bar{k}^{ab}A_b-\frac{2}{3}a_2k A^a}{ x}+\frac{4 a_1 n_{\mu}F^{\mu a}+2 a_2 D^a(n_{\mu}A^{\mu})}{ x}
  +..., \\
  % v10 J^{n}_S to J^{n}_A
&&J^{n}_A= \frac{-2 a_2 D_aA^a}{x}+..., \ \quad x \sim 0,  \label{chiralcurrentJn}
\end{eqnarray} 
where $n$ and $a$ denote respectively the normal and tangential directions, 
$D_a$ is the covariant derivative on the boundary.  Note that the above $A^a$ depends on $x$ generally. To see this, 
we take $ds^2=dx^2+\Big(h_{ab}-2x k_{ab}+O(x^2)\Big)dy^ady^b$ and $A_{\mu}=A_{0\ \mu}+x A_{1\ \mu }+O(x^2)$ 
in Gauss normal coordinate. Then we have $A^a=A_b(h^{ab}+2x k^{ab})+O(x^2)=A^{\ a}_0+ x A^{ \ a}_1+2x k^{ab} A_{0\ b}+O(x^2)$. 
To compare with the holographic calculations with constant axial vectors and extrinsic curvatures, we can set $ A_{1\ \mu}=n_{\mu}F^{\mu a}=D^a(n_{\mu}A^{\mu})=D_aA^a=0$ in (\ref{chiralcurrentJa},\ref{chiralcurrentJn}).


\begin{thebibliography}{00}





\bibitem{review}
  For a review, see for example,
  D.~E.~Kharzeev,
  ``The Chiral Magnetic Effect and Anomaly-Induced Transport,''
  Prog.\ Part.\ Nucl.\ Phys.\  {\bf 75} (2014) 133
  [arXiv:1312.3348 [hep-ph]];
  %%CITATION = doi:10.1016/j.ppnp.2014.01.002;%%
  %151 citations counted in INSPIRE as of 09 Jan 2018
  K.~Landsteiner,
  ``Notes on Anomaly Induced Transport,''
  Acta Phys.\ Polon.\ B {\bf 47} (2016) 2617
  [arXiv:1610.04413 [hep-th]].
  %%CITATION = doi:10.5506/APhysPolB.47.2617;%%

\bibitem{Vilenkin:1995um} 
  A.~Vilenkin,
  ``Parity nonconservation and neutrino transport in magnetic fields,''
  Astrophys.\ J.\  {\bf 451} (1995) 700.

  \bibitem{Vilenkin:1980fu}
  A.~Vilenkin,
  ``Equilibrium Parity Violating Current In A Magnetic Field,''
  Phys.\ Rev.\ D {\bf 22} (1980) 3080.
  %%CITATION = doi:10.1103/PhysRevD.22.3080;%%
  %227 citations counted in INSPIRE as of 09 Feb 2018

\bibitem{Giovannini:1997eg}
  M.~Giovannini and M.~E.~Shaposhnikov,
  ``Primordial hypermagnetic fields and triangle anomaly,''
  Phys.\ Rev.\ D {\bf 57} (1998) 2186
  [hep-ph/9710234].
  %%CITATION = doi:10.1103/PhysRevD.57.2186;%%
  %251 citations counted in INSPIRE as of 09 Feb 2018

  \bibitem{alekseev}
  A.Y. Alekseev, V. V. Cheianov, and J. Froehlich,
   Phys. Rev. Lett.{\bf  81} (1998) 3503 [cond-mat/9803346].
   
   


%\cite{Fukushima:2012vr}
\bibitem{Fukushima:2012vr}
  K.~Fukushima,
  ``Views of the Chiral Magnetic Effect,''
  Lect.\ Notes Phys.\  {\bf 871} (2013) 241
  [arXiv:1209.5064 [hep-ph]].
  %%CITATION = doi:10.1007/978-3-642-37305-3_9;%%
  %41 citations counted in INSPIRE as of 09 Feb 2018

  %\cite{Kharzeev:2007tn}
\bibitem{Kharzeev:2007tn}
  D.~Kharzeev and A.~Zhitnitsky,
  ``Charge separation induced by P-odd bubbles in QCD matter,''
  Nucl.\ Phys.\ A {\bf 797}, 67 (2007)
  [arXiv:0706.1026 [hep-ph]].
  %%CITATION = doi:10.1016/j.nuclphysa.2007.10.001;%%
  %331 citations counted in INSPIRE as of 09 Feb 2018

\bibitem{Erdmenger:2008rm}
  J.~Erdmenger, M.~Haack, M.~Kaminski and A.~Yarom,
  ``Fluid dynamics of R-charged black holes,''
  JHEP {\bf 0901} (2009) 055
  [arXiv:0809.2488 [hep-th]].
  %%CITATION = doi:10.1088/1126-6708/2009/01/055;%%
  %326 citations counted in INSPIRE as of 09 Feb 2018

\bibitem{Banerjee:2008th}
  N.~Banerjee, J.~Bhattacharya, S.~Bhattacharyya, S.~Dutta, R.~Loganayagam
  and P.~Surowka,
  ``Hydrodynamics from charged black branes,''
  JHEP {\bf 1101} (2011) 094
  [arXiv:0809.2596 [hep-th]].
  %%CITATION = doi:10.1007/JHEP01(2011)094;%%
  %311 citations counted in INSPIRE as of 09 Feb 2018

\bibitem{Son:2009tf}
  D.~T.~Son and P.~Surowka,
  ``Hydrodynamics with Triangle Anomalies,''
  Phys.\ Rev.\ Lett.\  {\bf 103} (2009) 191601
  [arXiv:0906.5044 [hep-th]].

\bibitem{Landsteiner:2011cp}
  K.~Landsteiner, E.~Megias and F.~Pena-Benitez,
  ``Gravitational Anomaly and Transport,''
  Phys.\ Rev.\ Lett.\  {\bf 107} (2011) 021601
  [arXiv:1103.5006 [hep-ph]].

%\cite{Golkar:2012kb}
\bibitem{Golkar:2012kb}
  S.~Golkar and D.~T.~Son,
  ``(Non)-renormalization of the chiral vortical effect coefficient,''
  JHEP {\bf 1502}, 169 (2015)
 % doi:10.1007/JHEP02(2015)169
  [arXiv:1207.5806 [hep-th]].
  %%CITATION = doi:10.1007/JHEP02(2015)169;%%
  %68 citations counted in INSPIRE as of 16 Mar 2018

%\cite{Jensen:2012kj}
\bibitem{Jensen:2012kj}
  K.~Jensen, R.~Loganayagam and A.~Yarom,
  ``Thermodynamics, gravitational anomalies and cones,''
  JHEP {\bf 1302}, 088 (2013)
 % doi:10.1007/JHEP02(2013)088
  [arXiv:1207.5824 [hep-th]].
  %%CITATION = doi:10.1007/JHEP02(2013)088;%%
  %124 citations counted in INSPIRE as of 16 Mar 2018
  
  
   \bibitem{Chernodub:2016lbo}
  M.~N.~Chernodub,
  %``Anomalous Transport Due to the Conformal Anomaly,''
  Phys.\ Rev.\ Lett.\  {\bf 117}, no. 14, 141601 (2016)
 % doi:10.1103/PhysRevLett.117.141601
  [arXiv:1603.07993 [hep-th]].
  %%CITATION = doi:10.1103/PhysRevLett.117.141601;%%
  %9 citations counted in INSPIRE as of 10 Mar 2019

  %\cite{Chernodub:2017jcp}/Users/mac/Downloads/Physical Review Journals - Publicity Instructions for Authors.pdf
\bibitem{Chernodub:2017jcp}
  M.~N.~Chernodub, A.~Cortijo and M.~A.~H.~Vozmediano,
  %``Generation of a Nernst Current from the Conformal Anomaly in Dirac and Weyl Semimetals,''
  Phys.\ Rev.\ Lett.\  {\bf 120}, no. 20, 206601 (2018)
%  doi:10.1103/PhysRevLett.120.206601
  [arXiv:1712.05386 [cond-mat.str-el]].
  %%CITATION = doi:10.1103/PhysRevLett.120.206601;%%
  %10 citations counted in INSPIRE as of 10 Mar 2019

  


  
  %\cite{Chu:2018fpx,Chu:2019rod,Miao:2017aba,Miao:2018dvm,Chernodub:2018ihb,Chernodub:2019blw,Ambrus:2019khr}
\bibitem{Chu:2018fpx}
C.~Chu and R.~Miao,
%``Boundary String Current & Weyl Anomaly in Six-dimensional Conformal Field Theory,''
JHEP \textbf{07}, 151 (2019)
%doi:10.1007/JHEP07(2019)151
[arXiv:1812.10273 [hep-th]].

  
  %\cite{Chu:2019rod}
\bibitem{Chu:2019rod}
C.~Chu,
%``Weyl Anomaly and Vacuum Magnetization Current of M5‐brane in Background Flux,''
Fortsch.\ Phys.\  \textbf{67}, no.8-9, 1910005 (2019)
%doi:10.1002/prop.201910005
[arXiv:1903.02817 [hep-th]].
  
%\cite{Miao:2017aba}
\bibitem{Miao:2017aba}
R.~Miao and C.~Chu,
%``Universality for Shape Dependence of Casimir Effects from Weyl Anomaly,''
JHEP \textbf{03}, 046 (2018)
%doi:10.1007/JHEP03(2018)046
[arXiv:1706.09652 [hep-th]].  
  
  %\cite{Miao:2018dvm}
\bibitem{Miao:2018dvm}
R.~Miao,
%``Casimir Effect, Weyl Anomaly and Displacement Operator in Boundary Conformal Field Theory,''
JHEP \textbf{07}, 098 (2019)
%doi:10.1007/JHEP07(2019)098
[arXiv:1808.05783 [hep-th]].

  %\cite{Chernodub:2018ihb}
\bibitem{Chernodub:2018ihb}
  M.~N.~Chernodub, V.~A.~Goy and A.~V.~Molochkov,
  %``Conformal magnetic effect at the edge: a numerical study in scalar QED,''
  Phys.\ Lett.\ B {\bf 789}, 556 (2019)
 % doi:10.1016/j.physletb.2019.01.003
  [arXiv:1811.05411 [hep-th]].
  %%CITATION = doi:10.1016/j.physletb.2019.01.003;%%
  %1 citations counted in INSPIRE as of 19 Apr 2019
  
  %\cite{Chernodub:2019blw}
\bibitem{Chernodub:2019blw}
M.~Chernodub and M.~A.~Vozmediano,
%``Direct measurement of a beta function and an indirect check of the Schwinger effect near the boundary in Dirac-Weyl semimetals,''
Phys.\ Rev.\ Research.\  \textbf{1}, 032002 (2019)
%doi:10.1103/PhysRevResearch.1.032002
[arXiv:1902.02694 [cond-mat.str-el]].

%\cite{Ambrus:2019khr}
\bibitem{Ambrus:2019khr}
V.~E.~Ambrus and M.~Chernodub,
%``Helical vortical effects, helical waves, and anomalies of Dirac fermions,''
[arXiv:1912.11034 [hep-th]].

%\cite{Zheng:2019xeu}
\bibitem{Zheng:2019xeu}
J.~Zheng, D.~Li, Y.~Zeng and R.~Miao,
%``Anomalous Current Due to Weyl Anomaly for Conformal Field Theory,''
Phys.\ Lett.\ B \textbf{797}, 134844 (2019)
%doi:10.1016/j.physletb.2019.134844
[arXiv:1904.07017 [hep-th]].



%\cite{Hu:2020puq}
\bibitem{Hu:2020puq}
P.~J.~Hu, Q.~L.~Hu and R.~X.~Miao,
%``Note on anomalous currents for a free theory,''
Phys. Rev. D \textbf{101}, no.12, 125010 (2020)
%doi:10.1103/PhysRevD.101.125010
[arXiv:2004.06924 [hep-th]].
%1 citations counted in INSPIRE as of 29 Jan 2021

%\cite{Kawaguchi:2020kce}
\bibitem{Kawaguchi:2020kce}
M.~Kawaguchi, S.~Matsuzaki and X.~G.~Huang,
%``Dynamic scale anomalous transport in QCD with electromagnetic background,''
JHEP \textbf{10}, 017 (2020)
%doi:10.1007/JHEP10(2020)017
[arXiv:2007.00915 [hep-ph]].
%2 citations counted in INSPIRE as of 29 Jan 2021


%\cite{Kurkov:2020jet,Kurkov:2018pjw,Fialkovsky:2019rum}
\bibitem{Kurkov:2020jet}
M.~Kurkov and D.~Vassilevich,
%``How many surface modes does one see on the boundary of a Dirac material?,''
Phys. Rev. Lett. \textbf{124}, no.17, 176802 (2020)
%doi:10.1103/PhysRevLett.124.176802
[arXiv:2002.06721 [hep-th]].
%3 citations counted in INSPIRE as of 29 Jan 2021


%\cite{Kurkov:2018pjw}
\bibitem{Kurkov:2018pjw}
M.~Kurkov and D.~Vassilevich,
%``Gravitational parity anomaly with and without boundaries,''
JHEP \textbf{03}, 072 (2018)
%doi:10.1007/JHEP03(2018)072
[arXiv:1801.02049 [hep-th]].


%\cite{Fialkovsky:2019rum,Vassilevich:2019mhl}
\bibitem{Fialkovsky:2019rum}
I.~Fialkovsky, M.~Kurkov and D.~Vassilevich,
%``Quantum Dirac fermions in a half-space and their interaction with an electromagnetic field,''
Phys.\ Rev.\ D \textbf{100}, no.4, 045026 (2019)
%doi:10.1103/PhysRevD.100.045026
[arXiv:1906.06704 [hep-th]].


  %\cite{Duff:1993wm}
\bibitem{Duff:1993wm} 
  M.~J.~Duff,
  %``Twenty years of the Weyl anomaly,''
  Class.\ Quant.\ Grav.\  {\bf 11}, 1387 (1994)
%  doi:10.1088/0264-9381/11/6/004
 % [hep-th/9308075].
  %%CITATION = doi:10.1088/0264-9381/11/6/004;%%
  %332 citations counted in INSPIRE as of 02 Nov 2017 
  
  %\cite{Chu:2018ksb}
\bibitem{Chu:2018ksb}
C.~S.~Chu and R.~X.~Miao,
%``Weyl Anomaly Induced Current in Boundary Quantum Field Theories,''
Phys. Rev. Lett. \textbf{121}, no.25, 251602 (2018)
%doi:10.1103/PhysRevLett.121.251602
[arXiv:1803.03068 [hep-th]].
%24 citations counted in INSPIRE as of 22 Sep 2022
  
  %\cite{Chu:2018ntx}
\bibitem{Chu:2018ntx}
C.~S.~Chu and R.~X.~Miao,
%``Anomalous Transport in Holographic Boundary Conformal Field Theories,''
JHEP \textbf{07}, 005 (2018)
%doi:10.1007/JHEP07(2018)005
[arXiv:1804.01648 [hep-th]].
%29 citations counted in INSPIRE as of 22 Sep 2022
  
 %\cite{Chu:2020mwx,Chu:2020gwq,Hu:2020puq,Kawaguchi:2020kce}
\bibitem{Chu:2020mwx}
C.~S.~Chu and R.~X.~Miao,
%``Fermion condensation induced by the Weyl anomaly,''
Phys. Rev. D \textbf{102}, no.4, 046011 (2020)
%doi:10.1103/PhysRevD.102.046011
[arXiv:2004.05780 [hep-th]].
%3 citations counted in INSPIRE as of 29 Jan 2021

%\cite{Chu:2020gwq}
\bibitem{Chu:2020gwq}
C.~S.~Chu and R.~X.~Miao,
%``Weyl Anomaly induced Fermi Condensation and Holography,''
JHEP \textbf{08}, 134 (2020)
%doi:10.1007/JHEP08(2020)134
[arXiv:2005.12975 [hep-th]].
%3 citations counted in INSPIRE as of 29 Jan 2021

%\cite{Chu:2022bhj}
\bibitem{Chu:2022bhj}
C.~S.~Chu and R.~X.~Miao,
%``Chiral Current induced by Torsional Weyl Anomaly in Dirac and Weyl Semimetals,''
[arXiv:2210.01382 [cond-mat.mes-hall]].
%0 citations counted in INSPIRE as of 06 Oct 2022
% v4

%3 citations counted in INSPIRE as of 29 Jan 2021


%\cite{Chodos:1974je}
\bibitem{Chodos:1974je} 
  A.~Chodos, R.~L.~Jaffe, K.~Johnson, C.~B.~Thorn and V.~F.~Weisskopf,
  %``A New Extended Model of Hadrons,''
  Phys.\ Rev.\ D {\bf 9}, 3471 (1974).
  %doi:10.1103/PhysRevD.9.3471
  %%CITATION = doi:10.1103/PhysRevD.9.3471;%%
  %2787 citations counted in INSPIRE as of 04 Dec 2019
  
  %\cite{Chodos:1974pn}
\bibitem{Chodos:1974pn} 
  A.~Chodos, R.~L.~Jaffe, K.~Johnson and C.~B.~Thorn,
  %``Baryon Structure in the Bag Theory,''
  Phys.\ Rev.\ D {\bf 10}, 2599 (1974).
 % doi:10.1103/PhysRevD.10.2599
  %%CITATION = doi:10.1103/PhysRevD.10.2599;%%
  %1110 citations counted in INSPIRE as of 04 Dec 2019
  
  %\cite{Billo:2016cpy}
\bibitem{Billo:2016cpy}
M.~Billo, V.~Goncalves, E.~Lauria and M.~Meineri,
%``Defects in conformal field theory,''
JHEP \textbf{04}, 091 (2016)
%doi:10.1007/JHEP04(2016)091
[arXiv:1601.02883 [hep-th]].
%202 citations counted in INSPIRE as of 07 Oct 2022



%\cite{Takayanagi:2011zk,Fujita:2011fp,Nozaki:2012qd}
\bibitem{Takayanagi:2011zk}
  T.~Takayanagi,
  %``Holographic Dual of BCFT,''
  Phys.\ Rev.\ Lett.\  {\bf 107} (2011) 101602
  [arXiv:1105.5165 [hep-th]].
  %%CITATION = doi:10.1103/PhysRevLett.107.101602;%%
  %68 citations counted in INSPIRE as of 14 Dec 2016
  
  
  %\cite{Fujita:2011fp}
\bibitem{Fujita:2011fp}
M.~Fujita, T.~Takayanagi and E.~Tonni,
%``Aspects of AdS/BCFT,''
JHEP \textbf{11}, 043 (2011)
%doi:10.1007/JHEP11(2011)043
[arXiv:1108.5152 [hep-th]].
%258 citations counted in INSPIRE as of 24 Sep 2022
  
 

%\cite{Miao:2018qkc}
\bibitem{Miao:2018qkc}
R.~X.~Miao,
%``Holographic BCFT with Dirichlet Boundary Condition,''
JHEP \textbf{02}, 025 (2019)
%doi:10.1007/JHEP02(2019)025
[arXiv:1806.10777 [hep-th]].
%14 citations counted in INSPIRE as of 30 Jan 2021

%\cite{Chu:2021mvq}
\bibitem{Chu:2021mvq}
C.~S.~Chu and R.~X.~Miao,
%``Conformal boundary condition and massive gravitons in AdS/BCFT,''
JHEP \textbf{01}, 084 (2022)
%doi:10.1007/JHEP01(2022)084
[arXiv:2110.03159 [hep-th]].
%8 citations counted in INSPIRE as of 24 Sep 2022



%\cite{Miao:2017aba}
\bibitem{Miao:2017aba}
R.~X.~Miao and C.~S.~Chu,
%``Universality for Shape Dependence of Casimir Effects from Weyl Anomaly,''
JHEP \textbf{03}, 046 (2018)
%doi:10.1007/JHEP03(2018)046
[arXiv:1706.09652 [hep-th]].
%30 citations counted in INSPIRE as of 24 Sep 2022

%\cite{Gynther:2010ed}
\bibitem{Gynther:2010ed}
A.~Gynther, K.~Landsteiner, F.~Pena-Benitez and A.~Rebhan,
%``Holographic Anomalous Conductivities and the Chiral Magnetic Effect,''
JHEP \textbf{02}, 110 (2011)
%doi:10.1007/JHEP02(2011)110
[arXiv:1005.2587 [hep-th]].
%138 citations counted in INSPIRE as of 24 Sep 2022

%\cite{Herzog:2015ioa,Fursaev:2015wpa}
\bibitem{Herzog:2015ioa}
  C.~P.~Herzog, K.~W.~Huang and K.~Jensen,
  %``Universal Entanglement and Boundary Geometry in Conformal Field Theory,''
  JHEP {\bf 1601}, 162 (2016)
  [arXiv:1510.00021 [hep-th]].
  %%CITATION = doi:10.1007/JHEP01(2016)162;%%
  %7 citations counted in INSPIRE as of 22 Dec 2016

%\cite{Fursaev:2015wpa}
\bibitem{Fursaev:2015wpa}
  D.~Fursaev,
  %``Conformal anomalies of CFT’s with boundaries,''
  JHEP {\bf 1512}, 112 (2015)
  [arXiv:1510.01427 [hep-th]].
  %%CITATION = doi:10.1007/JHEP12(2015)112;%%
  %7 citations counted in INSPIRE as of 22 Dec 2016




\end{thebibliography}
\end{document}